\documentclass[aps,pra,twocolumn,superscriptaddress,raggedbottom,showpacs,floatfix,longbibliography]{revtex4-1}

\usepackage{graphicx}
\usepackage{braket}                         
\usepackage{amssymb, amsmath, gensymb}
\usepackage{color}
\usepackage{footnote}

\def\nm{{\ {\rm nm}}}						% nm
\def\um{{\ \mu{\rm m}}}                                                         % um
\def\Rb87{^{87}\rm{Rb}}					% Rb 87
\def\Hz{{\ {\rm Hz}}}						% Hz
\def\kHz{{\ {\rm kHz}}}						% kHz
\def\GHz{{\ {\rm GHz}}}						% GHz
\def\ms{{\ {\rm ms}}}						% ms
\def\us{{\ \mu{\rm s}}}						% us
\def\nK{{\ {\rm nK}}}						% nK
\def\erfc{{\ {\rm erfc}}}                                                        %erfc

\def\Rb87{^{87}\rm{Rb}}					% Rb 87

\begin{document}

\title{Brownian motion of solitons in a Bose-Einstein condensate}
\author{L. M. Aycock}
\affiliation{Joint Quantum Institute, National Institute of Standards and Technology, and University of Maryland, Gaithersburg, Maryland, 20899, USA}
\affiliation{Cornell University, Ithaca, New York, 14850, USA}
\author{H. M. Hurst}
\affiliation{Joint Quantum Institute, National Institute of Standards and Technology, and University of Maryland, Gaithersburg, Maryland, 20899, USA}
\affiliation{Condensed Matter Theory Center, Department of Physics, University of Maryland, College Park, Maryland 20742, USA}
\author{D. K. Efimkin}
\affiliation{The Center for Complex Quantum Systems, The University of Texas at Austin, Austin, Texas 78712-1192, USA}
\author{D. Genkina}
\affiliation{Joint Quantum Institute, National Institute of Standards and Technology, and University of Maryland, Gaithersburg, Maryland, 20899, USA}
\author{H.-I Lu}
\affiliation{Joint Quantum Institute, National Institute of Standards and Technology, and University of Maryland, Gaithersburg, Maryland, 20899, USA}
\author{V. Galitski}
\affiliation{Joint Quantum Institute, National Institute of Standards and Technology, and University of Maryland, Gaithersburg, Maryland, 20899, USA}
\affiliation{Condensed Matter Theory Center, Department of Physics, University of Maryland, College Park, Maryland 20742, USA}
\author{I. B. Spielman}
\affiliation{Joint Quantum Institute, National Institute of Standards and Technology, and University of Maryland, Gaithersburg, Maryland, 20899, USA}
\date{\today}

\begin{abstract}
For the first time, we observed and controlled the Brownian motion of solitons. We launched solitonic excitations in highly elongated $\Rb87$ BECs and showed that a dilute background of impurity atoms in a different internal state dramatically affects the soliton. With no impurities and in one-dimension (1-D), these solitons would have an infinite lifetime, a consequence of integrability. In our experiment, the added impurities scatter off the much larger soliton, contributing to its Brownian motion and decreasing its lifetime.  We describe the soliton's diffusive behavior using a quasi-1-D scattering theory of impurity atoms interacting with a soliton, giving diffusion coefficients consistent with experiment.
\end{abstract}

\maketitle

\section{Introduction}

Solitons, spatially-localized, mobile excitations resulting from an interplay between nonlinearity and dispersion, are ubiquitous in physical systems from water channels and oceans to optical fibers and Bose-Einstein condensates (BECs).  From our pulse throbbing at our wrists to rapidly moving tsunamis, solitons appear naturally at a wide range of scales.  In non-linear optical fibers, solitons can travel long distances with applications to communication technology and potential for use in quantum switches and logic. Understanding how random processes contribute to the decay and the diffusion of solitons is essential to advancing these technologies.

We studied the diffusion and decay of solitons in the highly controlled quantum environment provided by atomic Bose-Einstein condensates (BECs), where density maxima can be stabilized by attractive interactions, i.e., bright solitons~\cite{Strecker2002, Khaykovich2002}; or as here, where density depletions can be stabilized by repulsive interactions, i.e., dark solitons such as kink solitons~\cite{Burger1999,Denschlag2000} and solitonic vortices~\cite{Ku2014}. By contaminating these BECs with small concentrations of impurity atoms, we quantitatively studied how random processes destabilize solitons.

Our BECs can be modeled by the one-dimensional (1-D) Gross-Pitaevski equation (GPE): an integrable nonlinear wave equation with soliton solutions as excitations above the ground state. For a homogeneous 1-D BEC of particles with mass $m_{\rm{Rb}}$ with density $\rho_0$, speed of sound $c$, and healing length $\xi=\hbar/\sqrt{2}m_{\rm{Rb}}c$, the dark soliton solutions
\begin{align}
\varphi(z,t)&=\sqrt{\rho_0}\left[i\frac{v_s}{c}+\frac{\xi}{\xi_s}\tanh\left(\frac{z-v_st}{\sqrt{2}\xi_s}\right)\right]
\label{Eqn:solWF}
\end{align}
are expressed in terms of time $t$, axial position $z$, soliton velocity $v_s$, and soliton width $\xi_s=\xi/\sqrt{1-(v_s/c)^2}$. Such dark solitons have a minimum density $\rho_0 (v_s/c)^2$ and a phase jump $-2\cos^{-1}(v_s/c)$ both dependent upon the soliton velocity $v_s$. These behave as classical objects with a negative inertial mass $m_s \approx -4\hbar \rho_0/c$, essentially the missing mass of the displaced atoms. The negative mass implies that increasing the soliton velocity reduces its kinetic energy, thus dissipation accelerates dark solitons~\cite{Muryshev2002}.  This can be seen from the soliton equation of motion 
\begin{align}
-|m_s|\ddot{z}(t) &= -\gamma\dot{z}(t) - \partial_zV + f(t),
\label{eqn:EOM}
\end{align}
where $-\gamma\dot{z}$ is the friction force and $V$ is the confining potential due to the mean-field effects of the condensate. The random Langevin force $f(t)$ has a white noise correlator $\langle f(t)f(t')\rangle = 2\gamma k_\mathrm{B}T\delta(t-t')$ where $T$ is temperature and $k_\mathrm{B}$ is Boltzmann's constant. The connection between the friction coefficient $\gamma$ and $f(t)$ derives from the same microscopic dynamics that yield the fluctuation-dissipation theorem for positive mass objects ---$f(t)$ is responsible for Brownian motion while $\gamma$ describes friction, but both have contributions from impurity atoms. 

Conventionally, the diffusion coefficient $D$ is inversely proportional to the friction coefficient: $D\propto 1/\gamma$. For negative mass objects, we show that the diffusion coefficient is instead proportional to the friction coefficient $D \propto \gamma$; this reflects that friction is an anti-damping force for negative mass objects. The interplay between friction and confinement drives diffusive behavior with linear-in-time variance in soliton position, ${\rm Var}(z) = D t$, the same Brownian motion present for positive mass objects.

Solitons are infinitely long-lived due to the integrability of the 1-D GPE. Integrability breaking is inherent in all physical systems, for example from the non-zero transverse extent of quasi-1-D systems.  Indeed the kink soliton in 3-D -- the direct analogue to the 1D GPE's dark soliton solution -- is only long-lived in highly elongated geometries~\cite{Becker2008, Weller2008, Hasegawa2003}, where integrability breaking is weak.  Cold atom experiments have profoundly advanced our understanding of soliton instability by controllably lifting integrability by tuning the dimensionality~\cite{Anderson2001, Ku2014}.  Here, we studied the further lifting of integrability by coupling solitons to a reservoir of impurities.

\begin{figure}[!htb]
\centering
   \includegraphics{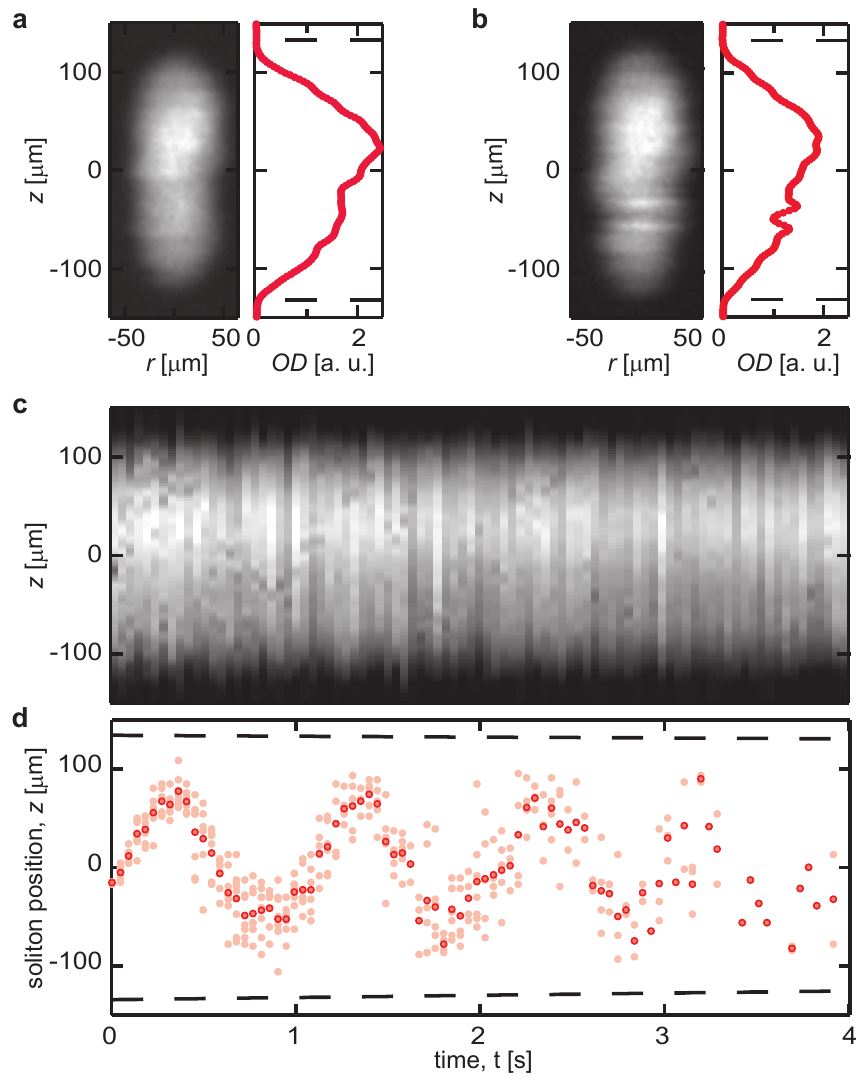}
   \caption{\textbf{Soliton oscillations.} \textbf{a}, An absorption image after a $19.3\ms$ TOF of an elongated condensate without a soliton and a longitudinal density distribution obtained by averaging over the remaining transverse direction. \textbf{b}, An absorption image and 1-D distribution at time $t=0.942\ \rm{s}$ with a soliton with $\approx 30\%$ imaged contrast. \textbf{c}, A subset of the data where each 1-D distribution is a unique realization of the experiment plotted versus time $t$. Notice a soliton was often absent at longer times. \textbf{d}, The axial position $z_i$ of the soliton (light pink) versus time $t$ for different realizations of the experiment. We repeated each measurement $8$ times. Dashed lines represent the edges of the elongated condensate. The dark markers represent the average soliton position $\langle z_i\rangle$  at each time $t$.}
   \label{solOsc}
\end{figure} 

\section{Experimental system}Our system~\cite{Lin2009a} consisted of an elongated $\Rb87$ BEC, confined in a nominally flat-bottomed time-averaged potential, created by spatially dithering one beam of our crossed dipole trap. We prepared $N=8(2) \times 10^{5}$ atoms~\footnote{In our system, number fluctuations increased at the lowest trap depth (see methods).} in the $\ket{F=1,m_{F}=0}$ internal state at $T=10(5)\ \nK$. Our system's $\approx250\um$ longitudinal extent was about $30$ times its transverse Thomas-Fermi diameter $2R_\perp$ set by the radial trap frequency $\omega_r=2 \pi\times 115(2) \Hz$ and chemical potential $\mu\approx h \times 1 \kHz$. We controllably introduced a uniform~\cite{Fang2015} gas of $N_{\rm I}$ impurity atoms in thermal equilibrium with our BECs using an rf pulse resonant with the $\ket{F=1,m_{F}=0}$ to $|F=1,m_F\!=\!+1\rangle$ transition prior to evaporation to degeneracy~\cite{Olf2015}. This gave impurity fractions $N_{\rm I}/N$ from $0$ to $0.062$ in our final BECs. 

We then launched long-lived solitonic excitations using a phase imprinting technique~\cite{Burger1999, Denschlag2000}.  Because our trap geometry had a finite transverse extent, quantified by the ratio $\mu/\hbar\omega_r\approx9$, planar kink solitons can be dynamically unstable and decay into 3-D excitations~\cite{Mateo2015}. Our soliton's initial velocity $v_s\approx0.3\ \rm{mm/s}$, roughly $1/5$ the 1-D speed of sound $c\approx1.4\ \rm{mm/s}$~\cite{Zaremba1998}, implies it is in an unstable regime~\cite{Muryshev2002}, where it will convert from a planar kink soliton to a nearly planar solitonic vortex.  For highly anisotropic geometries such as ours, the density profile of these two types of solitons is nearly the same -- as given by the 1D GPE -- reflecting that they become formally indistinguishable at large velocity~\cite{Mateo2015}.

\begin{figure}[!hb]
\centering
   \includegraphics{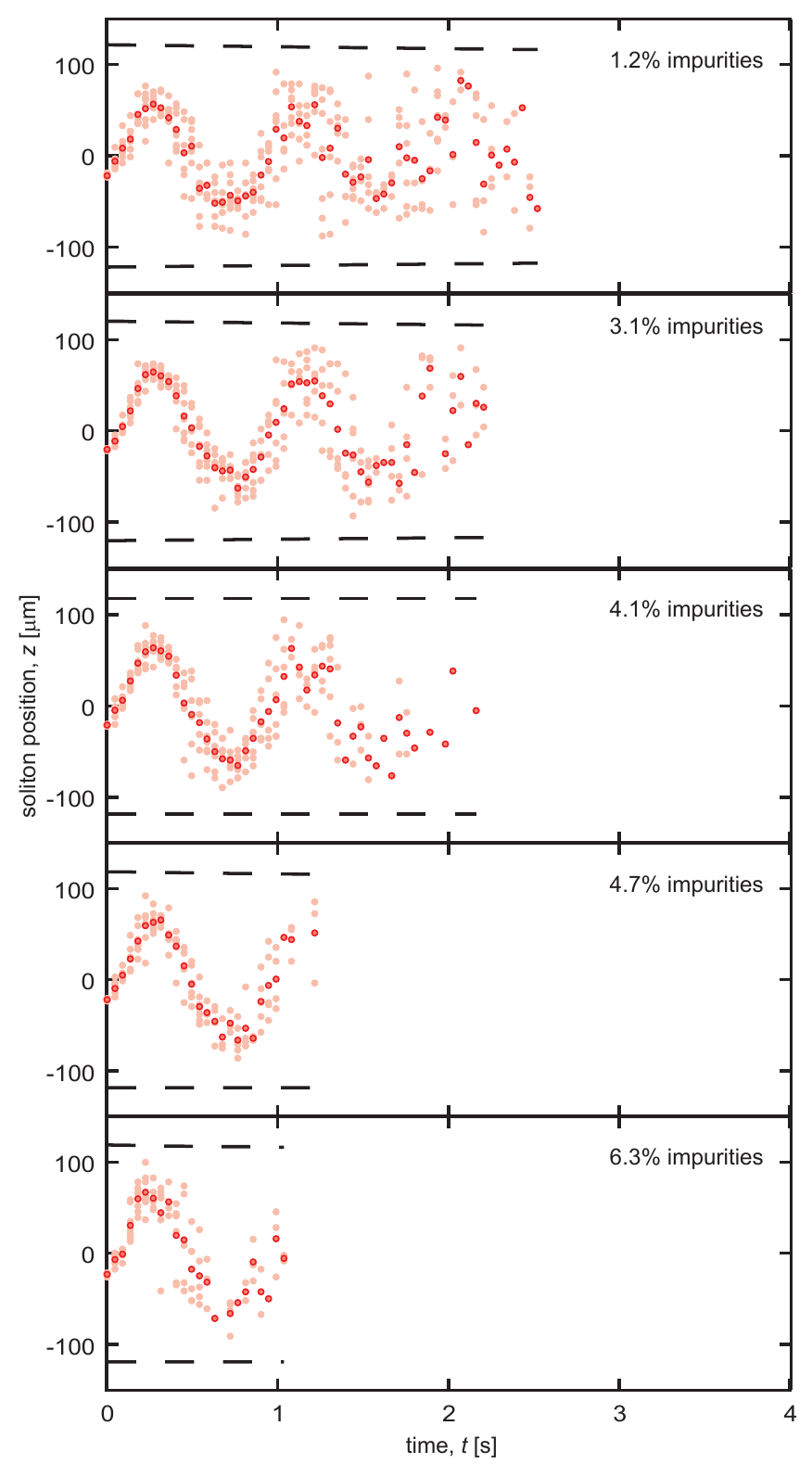}
   \caption{\textbf{Impact of impurities.} Here, we plot the position $z_i$ of the soliton (light pink) versus time $t$ after the phase imprint for different impurity levels. The dark pink markers represent the average position $\langle z_i\rangle$ for each time $t$. Dashed lines represent the endpoints of the condensate versus $t$.}
   \label{solWImp}
\end{figure}
 
We absorption-imaged our solitons after a sufficiently long time-of-flight (TOF) that their initial width $\xi_s\approx0.24\um$ expanded beyond our $\approx 2\um$ imaging resolution. Figure~\ref{solOsc}\textbf{a} shows our elongated BEC with no soliton present, and in contrast Fig.~\ref{solOsc}\textbf{b} displays a BEC with a soliton taken $0.947\ \rm{s}$ after its inception. The soliton is the easily identified density depletion sandwiched between two density enhancements. We quantitatively identified the soliton position as the minimum of the density depletion from 1-D distributions (right panel of Fig.~\ref{solOsc} \textbf{b}). Our phase imprinting process launched several excitations in addition to the soliton of interest. After a few hundred milliseconds, the additional excitations dissipated and the remaining soliton was identified. By backtracking the soliton trajectory, we were able to identify the soliton even at short times.

Figure~\ref{solOsc}\textbf{c} shows a series of 1-D distributions taken from time $t\approx0\ \rm{s}$ to $4\ \rm{s}$ after the phase imprint.  These images show three salient features: (1) the soliton underwent approximately sinusoidal oscillations, (2) the soliton was often absent at long times, and (3) there was significant scatter in the soliton position. Items (2) and (3) suggests that random processes were important to the soliton's behavior. The solitons' position $z_i$--when present--is represented by the light pink symbols in Fig.~\ref{solOsc}\textbf{d} and the darker pink symbols mark the average position $\langle z_i\rangle$ for each time $t$. 

\section{Coupling to impurities}
Having established a procedure for creating solitons, we turned to the impact of coupling to a reservoir of impurities, thus further breaking integrability. Figure~\ref{solWImp} displays the soliton position versus time for a range of impurity fractions. Adding impurities gave two dominant effects~\footnote{The soliton oscillation frequency was slightly shifted with impurities resulting from an unintentional change in the underlying optical potential. This change also slightly reduced the BECs longitudinal extent.}: further increasing the scatter in the soliton position $z$ and further decreasing the soliton lifetime. These effects manifested as a reduced fraction $f_s$ of images with a soliton present and an increase in the sample variance ${\rm Var}(z)=\sum \left(z_i-\langle z\rangle\right)^2/\left(M-1\right)$ computed using the number $M$ of measured positions $z_i$ at each time.  

\subsection{Reduced lifetime} The addition of impurities had a dramatic impact on the soliton lifetime. While we lack a quantitative model of the soliton's decay mechanism, there are several reasons to expect a finite lifetime. When dissipation is present, solitons accelerate to the speed of sound and disintegrate. Furthermore, numerical simulations show that in anharmonic traps solitons lose energy by phonon emission, accelerate, and ultimately decay~\cite{Parker2010}. All of these decay mechanisms can contribute to the soliton lifetime even absent impurities. 

\begin{figure*}[!htb]
\centering
   \includegraphics{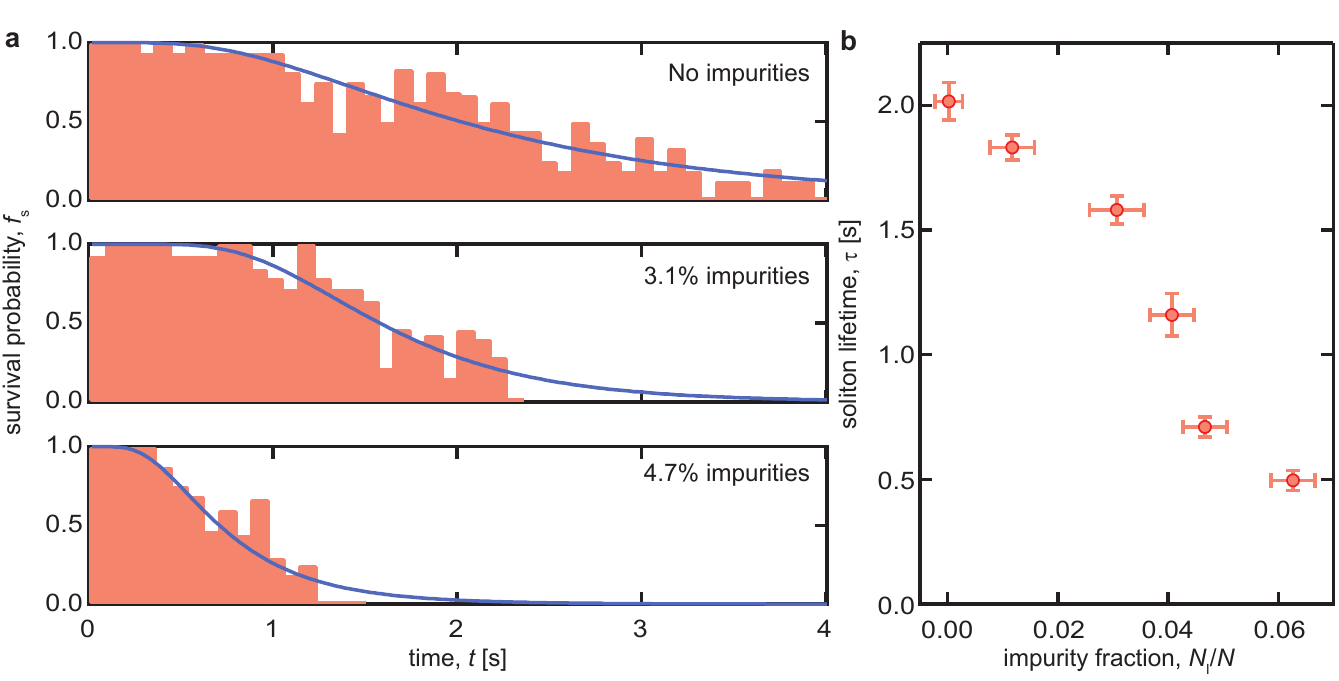}
   \caption{\textbf{Soliton lifetime in the presence of impurities.} \textbf{a}, Histograms of soliton occurrence probability $f_s$ versus time $t$ after phase imprint. The blue solid curves are fits to the lognormal based survival function from which we extract the lifetime $\tau$. For each impurity fraction, we stopped collecting data when $f_s$ fell below about $0.2$. \textbf{b}, Lifetime $\tau$ extracted from fit to the survival fraction $f_s$ versus impurity fraction $N_{\rm I}/N$.}
   \label{solLife}
\end{figure*}

The added impurities act as scatterers impinging on the soliton, further destabilizing it. This effect is captured in Fig.~\ref{solLife}\textbf{a}, showing the measured survival probability $f_s$ versus time for a range of impurity fractions. We fit to our data a model of the survival probability
\begin{align}
f_s(t) &= 1- \frac{1}{2}\erfc \left[\frac{-\ln(t/\tau)}{\sqrt{2}\sigma} \right],
\label{eqn:Survive}
\end{align}
essentially the integrated lognormal distribution of decay times, suitable for decay due to accumulated random processes~\cite{Cockburn2011}. The survival probability $f_s(t)$ has a characteristic width parameterized by $\sigma$ and reaches $1/2$ at time $\tau$ which we associate with the soliton lifetime. Figure~\ref{solLife}\textbf{b} shows the extracted lifetime $\tau$ versus impurity fraction $N_{\rm I}/N$, showing a monotonic decrease. Our maximum $N_{\rm I}/N$ gives a factor of four decrease in lifetime $\tau$. 

\subsection{Soliton diffusion} 
\begin{figure}[!ht]
\centering
   \includegraphics{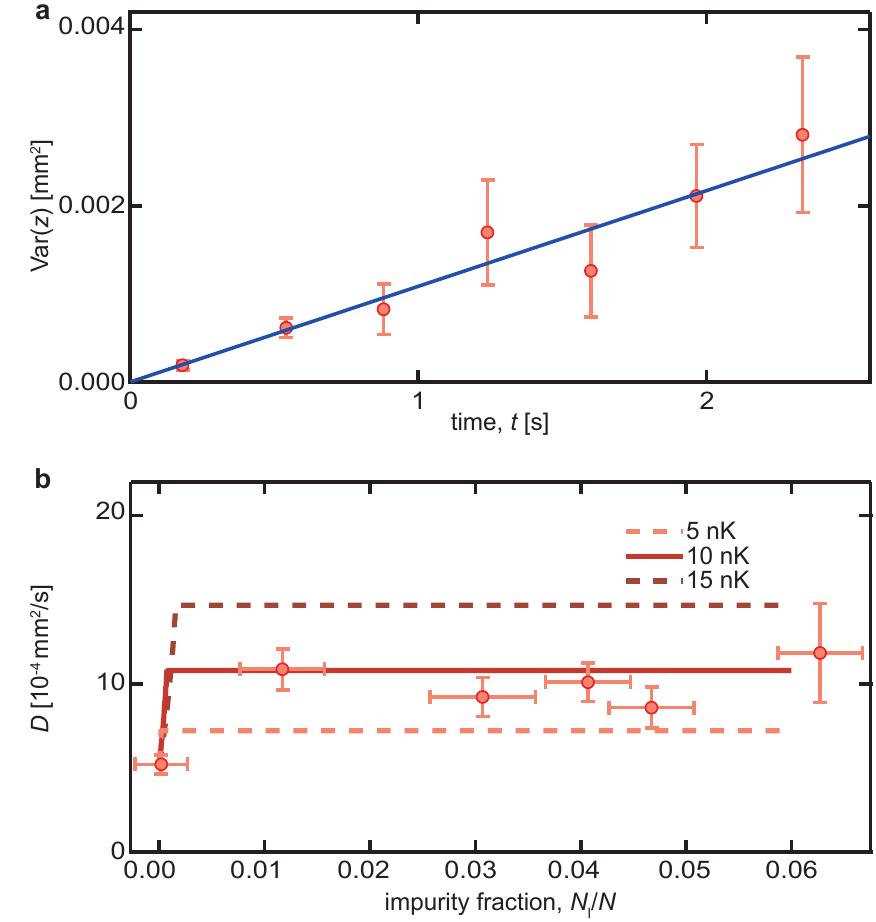}
   \caption{\textbf{Brownian diffusion constant dependence on impurities.} \textbf{a}, An example for the linear fit of ${\rm Var}(z)$ versus $t$ for 1.2\% impurities. Data is binned into 0.36s bins, the uncertainties are the sample standard deviation. \textbf{b}, The diffusion coefficient $D$ versus impurity fraction $N_{\rm I}/N$. The experimental results (markers) are extracted from the slope of a linear fit of the sample variance ${\rm Var}(z)$ versus time $t$. The uncertainty in $D$ is the uncertainty from that fit. See methods for explanation of uncertainty in $N_{\rm I}/N$. The theory curves (solid and dashed curves) plot the calculated $D$ for our measured temperature.}
   \label{solDiff}
\end{figure}

The second important consequence of adding impurities was an increased scatter in soliton position $z$, reminiscent of Brownian motion. Indeed, as shown in Fig.~\ref{solDiff}\textbf{a}, this scatter, quantified by ${\rm Var}(z)$, increased linearly with time. We obtained the diffusion coefficient $D$ as the slope from linear fits to these data and calculated $D$ using a quasi-1-D scattering theory. The energy of the infinitely long 1-D system is given by the GPE energy functional 
\begin{align}
&E\left[\varphi,\psi\right]=\nonumber \\
&\int \left( \frac{\hbar^2 |\nabla \varphi|^2}{2m_\mathrm{Rb}} + \frac{\hbar^2 |\nabla \psi|^2}{2m_{\rm{Rb}}} +  \frac{g}{2}|\varphi|^2|\varphi|^2 + \frac{g'}{2}|\varphi|^2|\psi|^2\right) dz,
\label{Eqn:Efunct}
\end{align}
describing the majority gas interacting with itself along with the impurities with interaction coefficients $g$ and $g'$, respectively. The fields $\varphi$ and $\psi$ denote the condensate and impurity wavefunctions. Since the impurities are very dilute, we do not include interactions between impurity atoms. A soliton [Eq.~(\ref{Eqn:solWF})] act as a supersymmetric P\"{o}schl-Teller~\cite{Poschl1933,Cooper1995} potential for the impurity atoms with exact solutions in terms of hypergeometric functions~\cite{Cevik2016}. Impurity scattering states with momentum $k_z$ in the rest frame of the soliton are described by the reflection coefficient 
\begin{align}
R(k_z) &= \frac{1-\cos(2\pi\lambda)}{\cosh(2\pi k_z \xi)-\cos(2\pi\lambda)},
\label{eqn:Rk}
\end{align}
where $\lambda(\lambda-1) = g'/g$. In $\Rb87$, we have $g \approx g'$, giving $\lambda \approx 1.5$. The scattering problem is fully characterized by $R(k_z)$ and the problem is reduced to that of a classical heavy object moving through a gas of lighter particles.

To understand soliton diffusion over many experimental runs we study their distribution function $f(t,z,v_s)$. We use a kinetic equation equivalent to Eq.~(\ref{eqn:EOM}) with a stochastic force due to elastic collisions with the impurity atoms and a harmonic confining potential $V(z) \approx -|m_s|\omega^2x^2/2$ where $\omega = \omega_{\rm trap}/\sqrt{2}$ is the effective frequency~\cite{Lifschitz1983,Konotop2004}. In the limit of small soliton velocity $(v_s/c)^2 \ll 1$, the time-dependent distribution function can be calculated exactly (see methods). The kinetic equation has no stable solutions: eventually all solitons accelerate and disappear.  However, the timescale for acceleration is set by $\Gamma^{-1} = |m_s|/\gamma$, is many seconds in our experiment. In the limit of $\Gamma t \ll 1$ and $\Gamma \ll \omega$, the variance in position grows linearly with time and diffusive behavior emerges, i.e. ${\rm Var}(z) \approx Dt$. We calculate the diffusion coefficient
\begin{align}
D&\approx \frac{\gamma + \gamma_0}{|m_s|\omega^2}\left(\frac{k_\mathrm{B}T}{|m_s|} + \frac{v_{\rm i}^2}{2}\right),
\label{eqn:Diff}
\end{align}
where $v_{\rm i}$ is the soliton's initial velocity. The offset $\gamma_0$ accounts for any diffusion present without impurities. The friction coefficient $\gamma$ is given by
\begin{equation}
\gamma = \frac{2\hbar}{k_\mathrm{B}T}\sum_{m,l}\int_{-\infty}^{\infty} \frac{dk_z}{2\pi}k_z^2\left|\frac{\partial \epsilon}{\partial k_z}\right|R(k_z)n(\epsilon)\left[1+n(\epsilon)\right],
\label{eqn:gamma}
\end{equation}
an extension of reference~\cite{Fedichev1999}. $\epsilon_{m,l}(k_z)=\hbar^2k_z^2/2m_\mathrm{Rb} + \hbar^2j_{m,l}^2/2m_{\rm{Rb}}R_\perp^2$ is the impurities' quasi-1D dispersion along with quantized states in the radial direction, described by Bessel functions. We account for radial confinement by summing over quantum numbers $m$ and $l$. $n(\epsilon)$ is the Bose-Einstein distribution for impurity atoms~\footnote{In our model, the condensed atoms do not contribute to the stochastic force underlying diffusion since they are all in the ground state. The impurity atoms are condensed for impurity fraction $\gtrsim0.2\%$, thus the number of thermal impurity atoms is constant, leading to constant friction coefficient.}. 

Figure \ref{solDiff}\textbf{b} plots $D$ measured experimentally (markers) and computed theoretically (curves, colored for different temperatures) as a function of $N_{\rm I}/N$. The theory provides rather accurate estimates of the experimentally observed diffusion coefficient, with a single fitting parameter given by $\gamma_0=5.32 \times 10^{-4}\ \rm{mm}^2/\rm{s}$.  $\gamma_0$ is set by the diffusion coefficient at $N_I/N=0$, where $D$ is suppressed in agreement with our theory.  Diffusion at zero impurity concentration could be due to a number of factors, including scattering of thermal phonons from the soliton as well as trap anharmonicity~\cite{Muryshev2002,Parker2010}. In our quasi-1-D system, the soliton is not reflectionless to phonons in the majority gas as it would be in 1-D. 

\section{Conclusion and outlook} 

Our data shows that added uncondensed impurity atoms contribute to soliton diffusion, however, the soliton-lifetime falls monotonically with increasing impurity fraction even when the additional impurities all enter the condensate.  We speculate that this might arise from two independent effects: (1) a static soliton forms a potential minimum for impurity atoms, implying that after some time impurities will congregate in these minima~\cite{Becker2008}, broadening and destabilizing the soliton; or (2) because the soliton moves in excess of the speed of sound for the impurity atoms, even condensed atoms can reflect from the moving soliton.  While this coherent reflection process would not add to diffusion, it would transfer momentum, thereby increasing the apparent damping coefficient and thereby reducing the soliton-lifetime.  This latter model predicts a reduction of lifetime qualitatively similar to, but quantitatively in excess of that observed in experiment. 

Solitons in spinor systems with impurity scatterers is an exciting playground for studying integrability breaking and diffusion of quasi-classical, negative-mass objects. Our observed reduction in soliton lifetime with increasing impurity fraction is in need of a quantitative theory. For the case of no impurities there is a further open question for both theory and experiment of whether friction and diffusion can be present even in the case of preserved integrability, for example due to non-Markovian effects, as was recently discovered for bright solitons~\cite{Efimkin2015}. Future experiments could study the impact of different types of impurities on soliton dynamics by introducing impurities of a different atomic mass. Lastly, mixtures with tunable interactions could freely tune the amount of impurity scattering, offering an additional way to change $D$. 

\section*{Acknowledgements}
We thank Martin Link and Stephen Eckel for carefully reading our manuscript.  We also benefited greatly from conversations with Joachim Brand.  This work was partially supported by the ARO's Atomtronics MURI, by the AFOSR's Quantum Matter MURI, NIST, and the NSF through the PFC at the JQI. HMH acknowledges additional fellowship support from the National Physical Science Consortium and NSA. VG acknowledges support from the Simons Foundation.

\section*{Author Contributions} L.M.A. configured the apparatus for this experiment and led the data acquisition and analysis effort and the manuscript preparation.  H.M.H. performed theoretical calculations and prepared the theory discussions in the manuscript. D.K.E. contributed to the development of the theoretical framework.  H.-I L. and D.G. contributed to all experimental aspects of this project and to the writing of the manuscript.  V.G. proposed the experiment concept and contributed to the development of the theory and to the writing of the manuscript.  I.B.S. developed the experimental concept, contributed to the understanding of experimental data and to the writing of the manuscript.
\section*{Materials and Methods}
\subsection*{BEC creation}
We created BECs in the optical potential formed by a pair of crossed horizontal laser beams of wavelength $\lambda=1064 \nm$~\cite{Lin2009a}. The beam traveling orthogonal to the elongated direction of the BEC was spatially dithered by modulating the frequency of an acoustic-optic modulator at a few hundred $\kHz$. This created an anharmonic, time-averaged potential.  To reach the extremely cold temperatures necessary to realize long lived solitons, we evaporated to the lowest dipole trap depth in which our technical stability allowed us to realize uniform BECs.

\subsection*{Temperature measurement}
We measured temperature below the majority atom's condensation temperature $T_c = 350\nK$ by removing the majority atoms and fitting the TOF expanded impurity atoms to a Maxwell-Boltzmann (MB) distribution~\cite{Olf2015}. Once the temperature was below $T_c$ for the impurity atoms, MB fits systematically under estimated the temperature. Fitting the small number of impurity atoms to a Bose distribution was technically challenging due to low signal-to-noise and the addition of another free parameter, the chemical potential. To limit the number of free parameters, we preformed a global fit on the different impurity fractions where we constrain the chemical potential $\mu$ to be negative. This provided an estimate of the temperature with large uncertainties. We found for our usual operating parameters and based on information from both temperature measurements, $T=10(5) \nK$.
 
\subsection*{Impurity characterization}
We use a Blackman enveloped rf pulse at a $\sim 9\ \rm{G}$ magnetic field to transfer the $\ket{F=1,m_{F}=0}$ atoms primarily to the $\ket{F=1,m_{F}=+1}$ internal state~\cite{Jimenez-Garcia2010a}. We varied the impurity fraction by tuning the rf amplitude. Even though the fraction of impurity atoms before evaporation determined the fraction after evaporation, they were not equal due to more effective evaporation of the minority spin state~\cite{Olf2015}. We characterized the impurity fraction through careful, calibrated absorption imaging with a Stern-Gerlach technique during TOF to measure the relative fraction of the impurity atoms after evaporation.

\subsection*{Soliton creation}
 We applied a phase shift to half of a condensate by imaging a back-lit, carefully-focused razor edge with light red detuned by $\approx6.8\GHz$ from the D$_2$ transition for $20\us$.
 
\subsection*{Scattering theory of impurities}
Minimizing Eq.~(\ref{Eqn:Efunct}) with respect to $\varphi^*$, $\psi^*$ gives the coupled equations of motion 
\begin{align}
i\hbar\partial_t\varphi(z,t) &= -\frac{\hbar^2}{2m_{\rm{Rb}}}\partial^2_{z}\varphi(z,t) + g|\varphi|^2\varphi + \frac{g'}{2}|\psi|^2\varphi \label{eqn:cond},\\ 
i\hbar\partial_t\psi(z,t) &= -\frac{\hbar^2}{2m_{\rm{Rb}}}\partial^2_{z}\psi(z,t) + \frac{g'}{2}|\varphi|^2\psi  \label{eqn:imp}.
\end{align}
In the experiment, we observed that the soliton remained stable for long times in the presence of impurities. Therefore we neglect the last term of Eq.~(\ref{eqn:cond}), giving the well known solitonic solution in Eq.~(\ref{Eqn:solWF}) of the main text. We seek a solution for the impurity wavefunction $\psi(z)$ in the soliton rest frame. In the radial direction the single particle wavefunctions are the usual Bessel functions for a particle in a cylindrical well. For $\psi(z)$ we combine Eq.~(\ref{Eqn:solWF}) and Eq.~(\ref{eqn:imp}) with $\psi(z,t) = e^{-iE_zt/\hbar}e^{im_\mathrm{Rb}v_sz'/\hbar}\psi(z')$. This gives a Schr\"{o}dinger equation with a P\"{o}schl-Teller potential~\cite{Poschl1933,Cevik2016}, 
\begin{align}
\frac{\partial^2\psi(z')}{\partial z'^2} + \left[\frac{\gamma_{s}^2\lambda(\lambda -1)}{\cosh^2(\gamma_{s} z')} +k_z^2\right]\psi(z') &= 0.
\label{eqn:psiz2}
\end{align}
The dimensionless parameters are $z'=(z-v_st)/\sqrt{2}\xi$, $k_z^2 = 4m_{\rm{Rb}}\xi^2/\hbar^2\left(E_z+m_{\rm{Rb}}v_s^2/2-g'\rho_0/2\right)$, $\lambda(\lambda -1 ) = 2m_{\rm{Rb}}\xi^2g'\rho_0/\hbar^2 = g'/g$, and $\gamma_{s}=\sqrt{1-(v_s/c)^2}$. $g$ and $g'$ are the effective 1-D interaction parameters after integrating over the transverse degrees of freedom in $\psi$ and $\varphi$. Since the transverse wavefunctions are different, in general $g'/g\lesssim 1$. However, $R(k_z)$ is periodic in $g'/g$ (through $\lambda$) and small variations in this parameter do not strongly affect the result. Solving for $\psi(z')$ and the scattering matrix then gives $R(k_z)$ Eq.~(\ref{eqn:Rk}) of the main text. For $\lambda \approx 1.5$, this potential also has a single, shallow bound state. Occupation of the bound state by an impurity atom can only occur through 3 body collisions (two impurity atoms and soliton), scenarios which we do not consider here.

\subsection*{Kinetic theory of the soliton}
In order to define a diffusion coefficient, we study the distribution function of many solitons, $f(t,z(t),v_s(t))$ (corresponding to many experimental runs). The distribution function of solitons follows a Boltzmann equation with a collision integral in Fokker-Planck form 
\begin{equation}
    \frac{df}{dt} = \frac{\partial}{\partial p}\left(Af+B\frac{\partial f}{\partial p}\right),
    \label{eqn:df}
\end{equation}
where $A$ and $B$ are the drift and diffusion transport coefficients and the left-hand side is a total time derivative. For $v_s\ll c$ we can write $A \approx \gamma v_s$ and $B \approx \gamma k_{\rm B} T$ where $v_s$ is the soliton velocity and $\gamma$ is the friction coefficient given in equation \eqref{eqn:gamma}. Finally, we write the soliton momentum as $p = -|m_s|v_s$~\cite{Fedichev1999}. The kinetic equation then takes the form
\begin{equation}
\frac{\partial f}{\partial t} + v_s\frac{\partial f}{\partial z} = \frac{\partial }{\partial v_s}\left(-\Gamma v_s f - \frac{\partial_zV}{|m_s|} f +\Gamma v^2_{\rm th}\frac{\partial f}{\partial v_s}\right), 
\label{eqn:Kinetic}
\end{equation}
where $\Gamma = \gamma/|m_s|$ and $v^2_{\rm th} = k_\mathrm{B}T/|m_s|$ is the thermal velocity. This equation can be solved analytically using the method of characteristics in the case of a harmonic potential $V(z) = -|m_s|\omega^2z^2/2$. The solution is the time-dependent distribution function $f(t,z,v_s)$, parametrized by functions $g_i(t,\omega)$ with Gaussian form 
\begin{align}
f(t,z,v_s) &= \frac{1}{2\pi\sqrt{4g_1g_3-g_2^2}}\exp\left\lbrace-\frac{1}{4g_1g_3-g_2^2}\left[g_1v_s^2+g_3z^2\right.\right.\nonumber \\
&+g_2v_sz +v_{\rm i}v_s(g_2g_4+2g_1g_5) + v_{\rm i}z(g_2g_5+2g_3g_4)\nonumber \\
&\left.\left.+v_{\rm i}^2(g_3g_4^2+g_1g_5^2+g_2g_4g_5)\right]\right\rbrace.
\label{eqn:RealSpacef}
\end{align}
Where $v_{\rm i}$ is the soliton initial velocity and functions $g_i(t,\omega)$ are given by 
\begin{align}
g_1(t,\omega) &= \frac{1+4\omega^2(e^t-1)-e^t\left[\cos(t\bar{\omega})+\bar{\omega}\sin(t\bar{\omega})\right]}{2\omega^2\bar{\omega}^2}\label{eqn:g1}\\
g_2(t,\omega) &=- \frac{2e^t}{\bar{\omega}^2}\left[1-\cos(t\bar{\omega})\right]\\
g_3(t,\omega) &= \frac{1+4\omega^2(e^t-1)+e^t\left[\bar{\omega}\sin(t\bar{\omega})-\cos(t\bar{\omega})\right]}{2\bar{\omega}^2}\\
g_4(t,\omega) &=- \frac{2e^{t/2}}{\bar{\omega}}\sin\left(\frac{t\bar{\omega}}{2}\right)\\
g_5(t,\omega) &= -\frac{e^{t/2}}{\bar{\omega}}\left[\sin\left(\frac{t\bar{\omega}}{2}\right) + \bar{\omega}\cos\left(\frac{t\bar{\omega}}{2}\right)\right]\label{eqn:g5}
\end{align}
Where we work in dimensionless units $t \rightarrow t/\Gamma$, $\omega \rightarrow \omega\Gamma$, $v_s\rightarrow v_{\rm th} v_s$, $z\rightarrow v_{\rm th}z/\Gamma$ and $\bar{\omega} = \sqrt{4\omega^2 -1}$. Equation \eqref{eqn:Kinetic} does not have a stable solution where $\partial f/\partial t \rightarrow 0$, due to the fact that the soliton is inherently unstable. The solution given in equation \eqref{eqn:RealSpacef} is valid for $v_s\ll c$. Finally, we calculate the variance in soliton position, ${\rm Var}(z)(t) = \int dv_s \int dz~z^2 f(t,z,v_s) = 2g_1 + v_{\rm i}^2g_4^2$, finding the exact expression (with restored units)
\begin{align}
{\rm Var}(z)(t) &= \frac{4v_{\rm th}^2(e^{\Gamma t}-1)}{4\omega^2-\Gamma^2} + \frac{4v_{i}^2e^{\Gamma t}}{4\omega^2-\Gamma^2}\sin^2\left(\frac{t\bar{\omega}}{2}\right)\nonumber\\
&+ \frac{v^2_{\rm th}\Gamma^2e^{\Gamma t}}{\omega^2(4\omega^2 -\Gamma^2)}\left[1-e^{\Gamma t}\left(\cos(t\bar{\omega}) + \frac{\bar{\omega}}{\Gamma}\sin(t\bar{\omega})\right)\right]
\end{align}
where $\bar{\omega} = \sqrt{4\omega^2-\Gamma^2}$. In the limits $\Gamma t \ll 1$, $\Gamma \ll \omega$, we find diffusive behavior ${\rm Var}(z) \approx D(t)t$ with the time-dependent diffusion coefficient
\begin{equation}
D(t) \approx \frac{v_{\rm th}^2\Gamma}{\omega^2} + \frac{v_{\rm i}^2\Gamma}{\omega^2}\sin^2\left(\frac{t\bar{\omega}}{2}\right)
\end{equation}
Setting $\sin^2(t\bar{\omega}/2) \approx 1/2$, we find the diffusion coefficient $D$ presented in equation \eqref{eqn:Diff}. We note that in the limit $\Gamma t\ll 1$, $\omega \rightarrow 0$, we have ${\rm Var}(z) \propto \Gamma t^3$ - the variance has no linear in $t$ dependence and the soliton undergoes ballistic motion, followed by exponential increase of ${\rm Var}(z)$ and soliton death~\footnote{Our trap was not strictly harmonic, however, the restoring force emerges for any trap with non-zero slope $\partial_zV \neq 0$, making our results qualitatively applicable to our experimental set.}. 

\bibliography{library}

\end{document}